\documentclass[english,aps,prl,superscriptaddress,twocolumn,showpacs,preprintnumbers,floatfix]{revtex4}
\usepackage[latin9]{inputenc}
\pdfoutput=1
\usepackage{nicefrac}
\usepackage{amsmath}
\usepackage{graphicx}
\usepackage{amssymb}

\makeatletter

\usepackage{amsfonts}
\usepackage{hyperref}
\usepackage{comment}
\usepackage{amsthm}
\usepackage{color}
\usepackage{units}

\makeatother

\usepackage{babel}
\makeatother

\begin{document}

\title{Parametric Normal-Mode Splitting in Cavity Optomechanics}

\author{J.M.~Dobrindt}

\affiliation{Max Planck\ Institut für Quantenoptik, D-85748 Garching, Germany.}

\author{I.~Wilson-Rae}

\affiliation{Technische Universität München, D-85748 Garching, Germany.}

\author{T.J.~Kippenberg}

\email[]{tjk@mpq.mpg.de}

\affiliation{Max Planck\ Institut für Quantenoptik, D-85748 Garching, Germany.}

\affiliation{Ecole Polytechnique Fédérale de Lausanne (EPFL), CH-1015 Lausanne, Switzerland.}

\keywords{Cooling, opto-mechanical coupling, radiation pressure,%
micro-mechanical oscillator, dynamical back-action, ground state}


\begin{abstract}
Recent experimental progress in cavity optomechanics has allowed cooling
of mesoscopic mechanical oscillators via dynamic backaction provided
by the parametric coupling to either an optical or an electrical resonator.
Here we analyze the occurrence of normal-mode splitting in backaction
cooling at high input power. We find that a hybridization of the oscillator's
motion with the fluctuations of the driving field occurs and leads
to a splitting of the mechanical and optical fluctuation spectra.
Moreover, we find that cooling experiences a classical limitation
through the cavity lifetime. 
\end{abstract}

\pacs{42.50.Wk, 03.65.Ta, 07.10.Cm}

\maketitle
\textit{Introduction}: Recently, cavity optomechanical systems that
parametrically couple a driven high-frequency mode to a high-Q, low-frequency
 mechanical mode have been subject to increasing investigation
\citep{kippenberg_cavity_2008}. They have been implemented in multiple
ways. Optomechanical systems have been demonstrated or proposed that
couple the mechanical motion to an optical field directly via radiation
pressure build up in a cavity \citep{braginsky_measurement_1977,kippenberg_analysis_2005,arcizet_radiation-pressure_2006,gigan_self-cooling_2006,schliesser_radiation_2006},
or indirectly via quantum dots \citep{wilson-rae_laser_2004} or ions
\citep{tian_coupled_2004}. On the other hand, in the electromechanical
domain, this has been realized or proposed using devices such as (superconducting)
single electron transistors \citep{naik_cooling_2006,lahaye_approaching_2004},
LC circuits \citep{brown_passive_2007}, a sapphire parametric transducer \citep{blair_high_1995}, 
Cooper pair boxes \citep{armour_entanglement_2002,Martin04},
or a stripline microwave resonator \citep{regal_measuring_2008}.
Importantly, the parametric coupling can not only be used for highly
sensitive readout of mechanical motion \citep{braginsky_measurement_1977}
but also by virtue of \textit{dynamical backaction} be used to cool
the mechanical oscillator. Indeed, recent progress has enabled 
the observation of radiation pressure dynamical backaction cooling \citep{gigan_self-cooling_2006,
  arcizet_radiation-pressure_2006,schliesser_radiation_2006} as predicted decades ago \citep{braginsky_measurement_1977,dykman_heating_1978}.
Enabled by this work, one emerging goal in this context is ground
state cooling, which may open up the possibility of studying nonclassical
states of motion or entanglement in mechanical objects \citep{tian_coupled_2004,mancini_entangling_2002,vitali07b}.
For both electro- and optomechanical systems, it has been shown that
ground state cooling is only possible in the resolved sideband regime
(RSB) where the mechanical resonance frequency exceeds the bandwidth
of the driving resonator \citep{wilson-rae_theory_2007,marquardt_quantum_2007}.
This result is analogous to the laser cooling of ions in the ``strong
binding'' regime \citep{wineland_laser_1979}. RSB cooling has recently been demonstrated
\citep{schliesser_resolved-sideband_2008,teufel_dynamical_2008}. \\%
\begin{figure}[b]
\centering\includegraphics[width=2.8in]{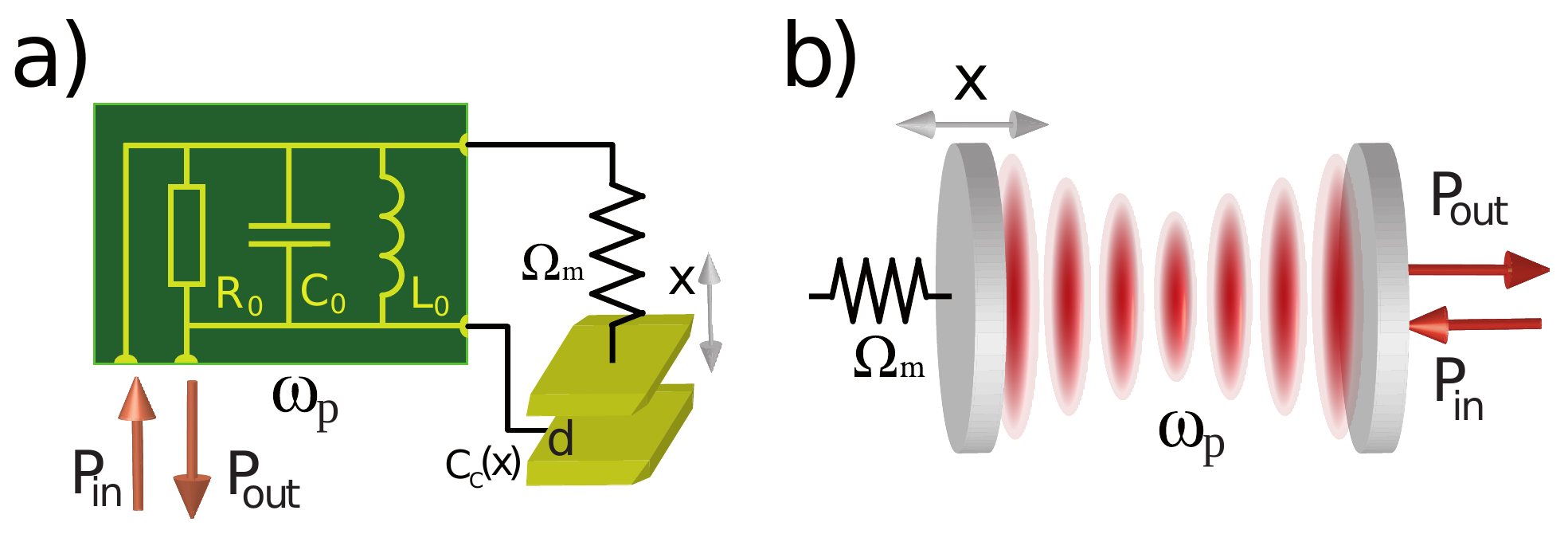} 
\caption{(a) Electromechanical realization of parametric coupling of a mechanical
oscillator to an \textit{LC} circuit, where the coupling is determined
by $\frac{d\omega_{p}}{dx}|_{x=0}=\frac{\omega_{p}C_{c}}{2dC_{tot}}|_{x=0}$
($C_{tot}$ is the  total capacitance). (b) Optomechanical realization of
parametric coupling of a mechanical oscillator to a Fabry-Perot optical
mode with $\frac{d\omega_{p}}{dx}|_{x=0}=-\frac{\omega_{p}}{L}|_{x=0}$
($L$ is the cavity length). \label{fig:1}}
\end{figure}
\indent Here we show that the cooling of mechanical oscillators in the RSB
regime at high driving power can entail the appearance of normal-mode
splitting (NMS). NMS --- the coupling of two degenerate modes with
energy exchange taking place on a timescale faster than the decoherence
of each mode --- is a phenomenon ubiquitous in both quantum and classical
physics. A prominent realization occurs when atoms are coupled to
a cavity field, which leads to the splitting of the cavity transmission
into a doublet \citep{thompson_observation_1992}. In addition to
atom-photon interactions, NMS also arises in exciton-photon and phonon-photon
interactions \citep{weisbuch_observation_1992}. NMS has also been
observed with ``artificial atoms'' in circuit QED \citep{wallraff_strong_2004}
and single quantum dot cavity QED \citep{Reithmaier_strong_2004}
settings. In these examples the NMS corresponds to a splitting in the 
energy spectrum of the coupled two-mode system which may be accessed 
via linear response. In contrast, the optomechanical NMS studied here 
involves driving two parametrically coupled non-degenerate modes out 
of equilibrium. Hence as will be discussed further below only in a ``shifted''  \citep{wilson-rae_theory_2007}
rotating-frame representation does the Hamiltonian become analogous to 
the one characterizing the aforementioned examples. Concomitantly, the 
splitting, rather than appearing directly in the cavity transmission, manifests 
itself in the \textit{fluctuation} spectra. This scenario is reminiscent of the single 
trapped ion realization of the Jaynes-Cummings model \citep{cirac93} 
with the role of the pseudospin now played by the optical (or electrical) mode. Since this 
type of normal-mode splitting occurs during RSB cooling, we analyze 
how the onset of NMS affects and limits cooling in the RSB regime.\\
\indent \textit{Theoretical model}: We start from the rotating-frame Hamiltonian
$H^{\prime}=-\hbar\Delta^{\prime}a_{p}^{\dagger}a_{p}+\hbar\Omega_{m}a_{m}^{\dagger}a_{m}+\hbar\eta\Omega_{m}a_{p}^{\dagger}a_{p}(a_{m}+a_{m}^{\dagger})+\hbar(s_{+}a_{p}+s_{+}^{*}a_{p}^{\dagger})$
which provides a unified treatment of both a coherently driven optical
and electrical resonator (frequency $\omega_{p}$) coupled to a mechanical
oscillator (frequency $\Omega_{m}\ll\omega_{p}$) via the dimensionless
parameter $\eta=(x_{0}/\Omega_{m})\frac{d\omega_{p}}{dx}|_{x=0}$.
Here $x_{0}=\sqrt{\hbar/2m_{eff}\Omega_{m}}$ is the zero point motion
of the mechanical mode, $m_{eff}$ its effective mass, $\Delta^{\prime}$
the detuning of the drive from $\omega_{p}$, and $a_{m}$ ($a_{p}$)
is the annihilation operator for the mechanical (optical or electrical)
mode. The dependence of the resonant frequency $\omega_{p}$ on the
mechanical oscillator's deflection $x$ determines the strength of
the coupling via $\frac{d\omega_{p}}{dx}|_{x=0}$ {[}cf.~Fig.~\ref{fig:1}].
The driving rate is given by $\left|s_{+}\right|=\sqrt{P/\hbar\omega_{p}\tau_{ex}}$,
where $P$ denotes the launched input power and $\tau_{ex}^{-1}$
is the external coupling rate.\newline
We derive the Heisenberg equations of motion for the canonical variables
and introduce noise operators $\xi_{m}(t)$ and $\xi_{p}(t)$ weighted
with the rates $\Gamma_{m}$ and $\kappa$ that characterize, respectively,
the dissipation of the mechanical and optical (or electrical) degree
of freedom. Subsequently, we shift the canonical variables to their
steady-state values (i.e.~$a_{p}\rightarrow\alpha+a_{p}$ and $a_{m}\rightarrow\beta+a_{m}$)
and linearize to obtain the following Heisenberg-Langevin equations
\citep{marquardt_quantum_2007,vitali07b,bhattacharya_trapping_2007}:
\begin{align}
\dot{a}_{p} & =\left(i\Delta-\frac{\kappa}{2}\right)a_{p}-i\frac{g_{m}}{2}\left(a_{m}+a_{m}^{\dagger}\right)+\sqrt{\kappa}\xi_{p}(t)\,,\label{eqn:lngvn}\\
\dot{a}_{m} & =\left(-i\Omega_{m}-\frac{\Gamma_{m}}{2}\right)a_{m}-i\frac{g_{m}}{2}\left(a_{p}+a_{p}^{\dagger}\right)+\sqrt{\Gamma_{m}}\xi_{m}(t).\nonumber \end{align}
 Here, $\Delta$ is the detuning with respect to the renormalized
resonance and $\Delta<0$ leads to cooling \citep{wilson-rae_theory_2007}.
The optomechanical coupling rate is given by $g_{m}=2\alpha\eta\Omega_{m}$,
which is positive by an appropriate choice for the
phase of $s_{+}$, and $|\alpha|^{2}$ gives the mean
resonator occupation number. In the case of the mechanical degree of freedom,
the rotating wave approximation in the coupling to its environment
implied by Eqs.~(\ref{eqn:lngvn}) is only warranted for high $Q$ values
(and small $g_{m}/\Omega_{m}$) \citep{vitali07b} --- conditions
that are satisfied in the parameter regime of interest for ground
state cooling. The latter also requires $\Gamma_{m}\ll\kappa$, which
we will assume throughout our treatment. Equations~(\ref{eqn:lngvn})
and their Hermitian conjugates constitute a system of four first-order
coupled operator equations, for which the Routh-Hurwitz criterion
implies that the system is only stable for $g_{m}<\sqrt{(\Delta^{2}+\kappa^{2}/4)\,\Omega_{m}/|\Delta|}\approx\Omega_{m}$
(if $\Omega_{m}\gg\kappa$ and $|\Delta|\approx\Omega_m$).\newline
Here, we follow a semi-classical theory by considering \textit{noncommuting}
noise operators for the input field, i.e., $\langle\xi_{p}(t)\rangle=0$,
$\langle\xi_{p}^{\dagger}(t^{\prime})\xi_{p}(t)\rangle=n_{p}\delta(t^{\prime}-t),\,\langle\xi_{p}(t^{\prime})\xi_{p}^{\dagger}(t)\rangle=\left(n_{p}+1\right)\delta(t^{\prime}-t)$,
and a \textit{classical} thermal noise input for the mechanical oscillator,
i.e.~$\langle\xi_{m}(t)\rangle=0$, $\langle\xi_{m}^{\dagger}(t^{\prime})\xi_{m}(t)\rangle=\langle\xi_{m}(t^{\prime})\xi_{m}^{\dagger}(t)\rangle=n_{m}\delta(t^{\prime}-t)$,
in Eqs.~(\ref{eqn:lngvn}). The quantities $n_{m}$ and $n_{p}$
are the equilibrium occupation numbers for the mechanical and optical
(or electrical) oscillators, respectively. We transform to the quadratures (i.e.,~$x/x_{0}=a_{m}+a_{m}^{\dagger}$)
and solve the Langevin equations in Fourier space \citep{marquardt_quantum_2007}.
Thus we recover a steady-state displacement spectrum \citep{bhattacharya_trapping_2007}
given (for $n_{p}=0$) by $S_{x}(\omega)=\frac{x_{0}^{2}}{2\pi}\Omega_{m}^{2}|\chi(\omega)|^{2}\left[\Gamma_{m}n_{m}-\frac{\Delta^{2}+\omega^{2}+\kappa^{2}/4}{2\Delta\Omega_{m}}\Gamma_{s}(\omega)\right]$ with \\
\begin{align}
\chi^{-1}(\omega) & =\Omega_{m}^{2}+2\Omega_{m}\Omega_{s}(\omega)-\omega^{2}-i\omega\left[\Gamma_{m}+\Gamma_{s}(\omega)\right]\label{q:shift}\\
\Omega_{s}(\omega) & =\frac{g_{m}^{2}}{4}\,\left[\frac{\omega+\Delta}{(\omega+\Delta)^{2}+\kappa^{2}/4}-\frac{\omega-\Delta}{(\omega-\Delta)^{2}+\kappa^{2}/4}\right]\nonumber \\
\Gamma_{s}(\omega) & =\frac{g_{m}^{2}}{4\omega}\,\left[\frac{\Omega_{m}\kappa}{(\omega+\Delta)^{2}+\kappa^{2}/4}-\frac{\Omega_{m}\kappa}{(\omega-\Delta)^{2}+\kappa^{2}/4}\right]\,.\nonumber \end{align}
 This spectrum is characterized by a mechanical susceptibility $\chi(\omega)$
that is driven by thermal noise ($\propto n_{m}$) and by the quantum
fluctuations of the radiation pressure (quantum backaction).
In linear cooling theory the susceptibility is
approximated by evaluating the terms $\Gamma_{s}(\omega)$ and $\Omega_{s}(\omega)$
at the (bare) mechanical frequency \citep{arcizet_radiation-pressure_2006,schliesser_radiation_2006,corbitt_optical_2007}.
Then $\Gamma_{s}(\Omega_{m})$ coincides with
the cooling rate and is linear in the input power ($g_{m}^{2}\propto P$).\\
\indent \textit{Parametric NMS}: The above approximation is only adequate
for weak driving such that $g_{m}\ll\kappa$ \citep{wilson-rae_theory_2007,marquardt_quantum_2007}.
To obtain an understanding of the mechanical susceptibility beyond
this linear regime, we return to the linearized Heisenberg-Langevin
equations~(\ref{eqn:lngvn}) and calculate the corresponding eigenfrequencies
that determine the dynamics of the system. Though there exists an
analytical solution, it is rather opaque and does not provide physical
insight, so that we will use instead an approximation scheme appropriate
for the parameter regime relevant for the observation of NMS and to
attain ground state cooling. Along these lines, we focus in the following
on: ($i$) the RSB regime ($\kappa\lesssim\Omega_{m}/2$) necessary
for ground state cooling \citep{wilson-rae_theory_2007,marquardt_quantum_2007,schliesser_resolved-sideband_2008},
($ii$) optomechanical coupling $g_{m}\lesssim\Omega_{m}/2$, and
($iii$) $\delta^{2}\ll\Omega_{m}^{2}$ ($\delta\equiv-\Delta-\Omega_{m}$, the frequency detuning from the lower sideband).
\begin{figure}[b]
\centering\includegraphics[width=3.2in]{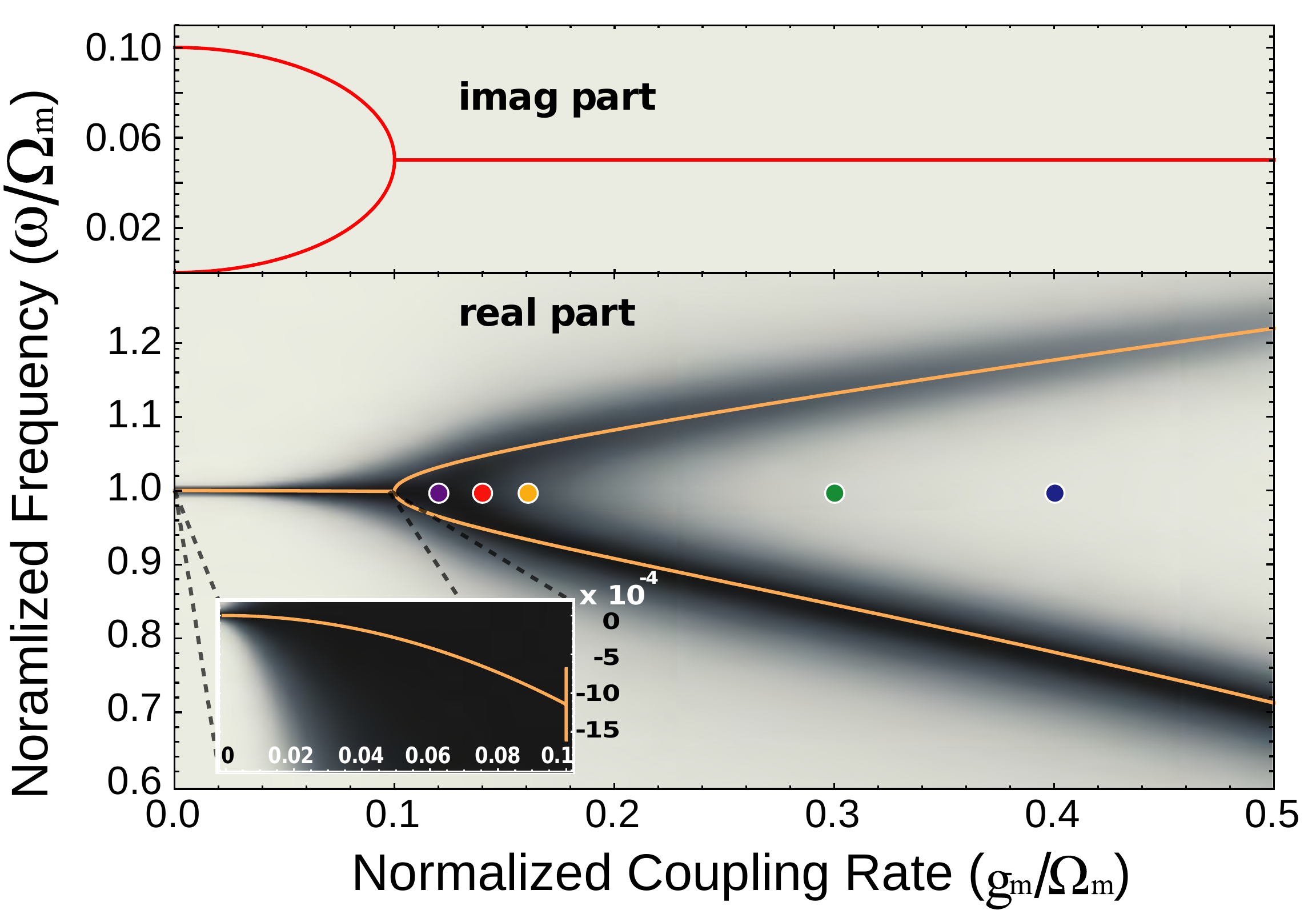} 
\caption{Real and imaginary parts of the eigenvalues {[}cf. Eqs. (\ref{eqn:gnvls0})
and (\ref{eqn:gnvls2})] of the linearized cooling problem 
corresponding, respectively, to the eigenfrequency and mode damping for $\Delta=-\Omega_{m}$
and $\kappa$/$\Omega_{m}=0.2$. The inset magnifies the resonance
shift before the mode splitting. The real part is underlaid with the
normalized classical displacement spectrum (contribution $\propto n_{m}$)
{[}cf.~Eq.~\ref{q:shift}], thereof sample curves are highlighted
in Fig.~\ref{fig:splitb} for the rates marked by the dots. \label{fig:splita}}
\end{figure}
In the shifted representation corresponding to Eqs.~(\ref{eqn:lngvn})
\citep{wilson-rae_theory_2007} the relevant part of the parametric
interaction in Hamiltonian $H'$ is described by an effective
dipole-like interaction term, i.e. $\hbar\eta\Omega_{m}a_{p}^{\dagger}a_{p}(a_{m}+a_{m}^{\dagger})\to\frac{\hbar g_{m}}{2}(a_{p}+a_{p}^{\dagger})(a_{m}+a_{m}^{\dagger})$
after neglecting the nonlinear term. This interaction term is analogous
to the Jaynes-Cummings setting (with $a_{p}\to\sigma^{-}$) and naturally
leads to resonance splitting when the modes have matching frequencies.
The off-resonant counterrotating terms (CRT) $\propto a_{p}^{\dag}a_{m}^{\dag}$,
$a_{p}^{\vphantom{\dag}}a_{m}^{\vphantom{\dag}}$ induce a small frequency
shift analogous to the Bloch-Siegert shift in atomic physics \citep{Bloch_magnetic_1940}.
These CRT terms, which are responsible for the mixing between the
creation and annihilation operators in the Heisenberg-Langevin equations ~(\ref{eqn:lngvn}),
can be treated in perturbation theory within the parameter range defined
by ($i$)-($iii$). The first nonvanishing order in this perturbative
expansion is quadratic in the CRT and yields a correction to the decoupled
eigenvalues $\omega_{\pm}\approx\omega_{\pm}^{(0)}+\omega_{\pm}^{(2)}$
{[}note that we take $\Gamma_{m}=0$ in $\omega_{\pm}^{(2)}$]: \\
\begin{align}
\omega_{\pm}^{(0)}= & \,\,\Omega_{m}+\frac{\delta}{2}-i\frac{\kappa+\Gamma_{m}}{4}\nonumber \\
 & \pm\frac{1}{2}\sqrt{g_{m}^{2}-(\kappa/2-\Gamma_{m}/2+i\delta)^{2}}\,,\label{eqn:gnvls0}\\
\omega_{\pm}^{(2)}\approx & -\frac{g_{m}^{2}/4}{2\Omega_{m}+\delta\pm\sqrt{g_m^2-(\kappa/2+i\delta)^2)}}\,.\label{eqn:gnvls2}\end{align}
 Naturally, there is another pair of eigenfrequencies given by $-\omega_{\pm}^{\ast}$.
In Fig.~\ref{fig:splita}, the real and imaginary parts of the eigenvalues
are plotted. The inset shows the frequency shift ($\omega_{\pm}^{(2)}$)
due to the CRT. If we choose the value $\delta=0$ (i.e., $\Delta = -\Omega_m$) relevant for $\kappa\ll\Omega_{m}$
(see below) and neglect $\Gamma_{m}$, the square root term of $\omega_{\pm}^{(0)}$
leads to two regimes. While for $g_{m}<\kappa/2$ the term is fully
imaginary and modifies the decay rate of the modes, for $g_{m}>\kappa/2$
it becomes real instead and the real parts of the eigenfrequencies
exhibit the splitting that signals NMS (Fig. \ref{fig:splitb}). The
latter is associated to a mixing between the mechanical mode and the
\textit{fluctuation} around the steady-state of the resonator field. Classically,
this \textit{fluctuation} can be understood as a beat of the pump
photons with the photons scattered on resonance which leads to oscillations
with frequency $|\Delta|$ in the intensity time-averaged over $2\pi/\omega_{p}$.
For $\kappa^{2}/4\!\!\ll\!\! g_{m}^{2}$ the splitting ($\approx\!\! g_{m}$)
is proportional to the square root of the mean cavity photon number ($\alpha^2$). This
is analogous to NMS in atomic physics where the splitting of the cavity
resonance is proportional to the square root of the number of atoms
coupled to the cavity mode \citep{thompson_observation_1992}. When detecting the phase fluctuations
in the transmitted light with a homodyne detection scheme, the signal at $\Omega_{m}$
splits {[}cf.~Fig.~\ref{fig:splitb}b], but the (suppressed) scattered
light at the carrier frequency exhibits no splitting. It is important
to note that the splitting in the displacement spectrum is not observed
unless $g_{m}>\kappa/\sqrt{2}$ due to the finite width of the peaks. 
Due to the requirements on the cavity bandwidth and the detuning,
the parameter regime in which NMS may appear implies cooling. In turn,
for a positive detuning (which entails amplification) the
observation of NMS is prevented by the onset of the parametric instability
\citep{kippenberg_analysis_2005}. Therefore, a discussion of NMS
cannot be decoupled from an analysis of the associated cooling. We
also show below that the CRT in the interaction lead to the quantum
limit of backaction cooling \citep{marquardt_quantum_2007,wilson-rae_theory_2007}.\\
\begin{figure}[b]
\centering\includegraphics[width=3.2in]{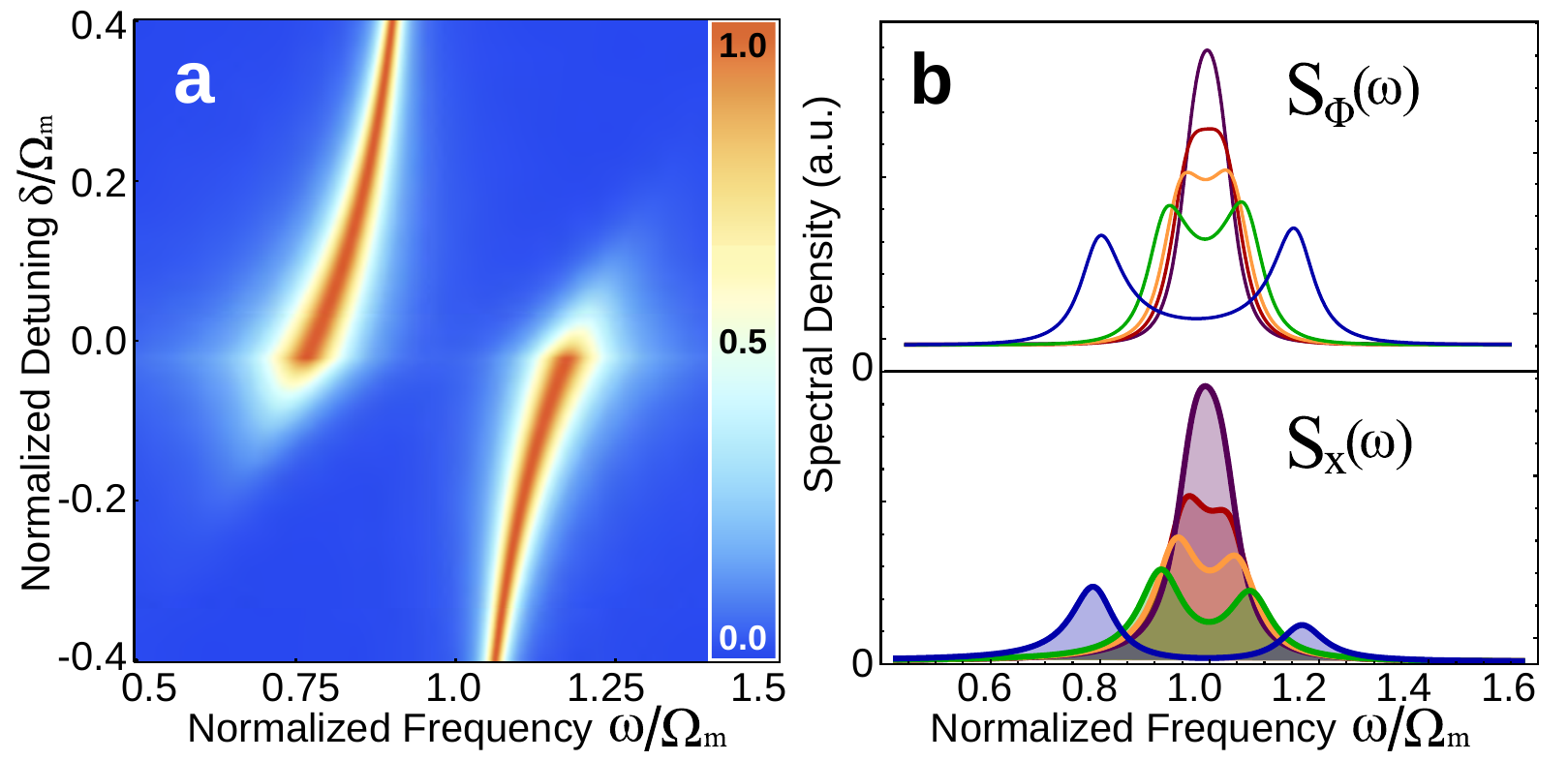}
\caption{(a) Normalized logarithm of the classical displacement spectrum (contribution
$\propto n_{m}$) {[}cf.~Eq.~\ref{q:shift}] as a function of the
normalized detuning for $g_{m}/\Omega_{m}=0.4$ and $\kappa/\Omega_{m}=0.2$.
(b) Typical displacement spectra for coupling rates that are represented
by the dots in Fig. \ref{fig:splita}. The solid curves above correspond
to the phase spectral density $S_{\Phi}(\omega)$ measured in homodyne
detection. \label{fig:splitb}}
\end{figure}
\indent \textit{Effect of NMS on backaction cooling}: We now use the approximate
eigenfrequencies to perform contour integration on the normal ordered
mechanical spectrum in order to obtain the final occupancy
of the mechanical oscillator $n_{f}=\left.\langle a_{m}^{\dagger}(\tau)a_{m}(0)\rangle\right|_{\tau=0}$.
In this treatment we take both the thermal and the vacuum
noise of the driving resonator into account. A finite value for $n_{p}$
may be relevant for electromechanical systems \citep{brown_passive_2007, regal_measuring_2008}.
Within our approximation scheme we can introduce a formal parameter
that tags the CRT terms and expand $n_{f}$ in its powers. To zeroth
order the poles are determined by the approximate eigenfrequencies
$\omega_{\pm}^{(0)},-\omega_{\pm}^{(0)*}$ given in Eqs.~(\ref{eqn:gnvls0}),
and it is straightforward to evaluate $n_{f}^{(0)}$ (including $\Gamma_{m}$).
To second order we use instead the poles $\omega_{\pm}^{(0)}+\omega_{\pm}^{(2)},-\omega_{\pm}^{(0)*}-\omega_{\pm}^{(2)*}$.
Subsequently, $n_{f}^{(2)}$ is expanded in the \textit{small} parameters
$g_{m}/\Omega_{m}$, $\kappa/\Omega_{m}$, and $|\delta|/\Omega_{m}$
up to second order with $\Gamma_{m}\to0$. Both $n_{f}^{(0)}$ and
$n_{f}^{(2)}$ do not contain terms linear in $\delta$, allowing one to directly minimize 
the result with respect to $\delta$ by setting $\delta\to0$. This yields \begin{eqnarray}
n_{f}^{(0)} & = & n_{m}\frac{\Gamma_{m}}{\kappa}\frac{g_{m}^{2}+\kappa^{2}}{g_{m}^{2}+\Gamma_{m}\kappa}+\frac{g_{m}^{2}}{g_{m}^{2}+\Gamma_{m}\kappa}n_{p}\,,\nonumber \\
n_{f}^{(2)} & = & n_{m}\frac{\Gamma_{m}}{\kappa}\frac{g_m^{2}}{4\Omega_{m}^{2}}+\left(n_{p}+\frac{1}{2}\right)\frac{\kappa^{2}+2g_{m}^{2}}{8\Omega_{m}^{2}}\,.\label{eqn:nf}\end{eqnarray}
The final occupancy $n_{f}=n_{f}^{(0)}+n_{f}^{(2)}$ consists of three
contributions. One is proportional to the occupancy of the thermal
bath $n_{m}$ and displays linear cooling for $\Gamma_{m}\ll g_{m}\ll\kappa$,
i.e.,~$n_{f}\approx\frac{\Gamma_{m}}{g_{m}^{2}/\kappa}n_{m}$. When
$g_{m}$ approaches $\kappa$, deviations from the linear cooling
regime become apparent. Indeed, the final occupancy is \textit{always
limited} by $n_{f}\gtrsim n_{m}\frac{\Gamma_{m}}{\kappa}$, which implies
that the largest temperature reduction is bound by the cavity decay
rate $\kappa$ \footnote{Note that $n_{f}^{(0)}$ follows from the classical rate equations
for two resonant oscillators (frequency $\Omega_{m}$) connected,
respectively, to two reservoirs at temperatures $T_{m}$ and $T_{\mathrm{eff}}=T_{m}\Omega_{m}/\omega_{p}$
via rates $\Gamma_{m}$ and $\kappa$ ($\Gamma_{m}\ll\kappa$), and
coupled via heat diffusion with a rate $g_{m}^{2}/\kappa$. In this
picture, the deviation from linear cooling corresponds to heat diffusion
from the cavity to the mechanical oscillator.}. This is equivalent to the condition $Q_{m}>n_{m}\frac{\Omega_{m}}{\kappa}$
for ground state cooling.  It is noted that operation in the deeply RSB regime is advantageous 
to avoid photon-induced heating \citep{schliesser_resolved-sideband_2008}, 
entailing that the condition on the mechanical $Q$ is therefore more stringent.
A second contribution is proportional to the finite occupancy of the
driving circuit ($n_{p}$) and corresponds to heating from thermal
noise in its input. It implies that it is impossible to cool below
the equilibrium occupation of the resonator. If we assume that
the mechanical and electromagnetic baths are at the same temperature
$T_{m}$, it entails $n_{f}\geq n_{m}\frac{\Omega_{m}}{\omega_{p}}$. Last, there is a term in $n_{f}^{(2)}$
that is temperature-independent and corresponds to heating from quantum backaction
noise. This term determines the quantum limit to the final occupancy
and agrees with Refs.$\,$\citep{marquardt_quantum_2007,wilson-rae_theory_2007}. 
Interestingly, in the present analysis the quantum limit arises from
the CRT. We note that the trade-off  between the quantum 
limit and the cavity bandwidth limitation leads to an optimal value for $\kappa$.
Consistent results are obtained with a covariance matrix
approach \citep{wilson-rae_cavity_2008}. \newline
Finally, we consider appreciable cooling {[}$n_{f}\ll n_{m}$ so that
we can take $\Gamma_{m}\to0$ in the denominator of Eqs.~(\ref{eqn:nf})]
and optimize $n_{f}^{(0)}+n_{f}^{(2)}$ with respect to $g_{m}$,
which yields \begin{equation}
n_{\text{opt}}\approx n_{m}\frac{\Gamma_{m}}{\kappa}+n_{p}+\frac{\kappa^{2}}{16\Omega_{m}^{2}}+\sqrt{\frac{n_{m}\Gamma_{m}\kappa(n_{p}+\nicefrac{1}{2})}{\Omega_{m}^{2}}}\label{eqn:nopt}\end{equation}
 for $g_{\text{opt}}=\sqrt[4]{4n_{m}\Gamma_{m}\kappa\Omega_{m}^{2}/[n_{p}+\nicefrac{1}{2}+n_m\Gamma_m/\kappa])}$.
In the ground state cooling regime, the first three terms of Eq.~(\ref{eqn:nopt})
always give the correct order of magnitude. Thus, a comparison of
$g_{\text{opt}}$ with the condition $g_{m}>\kappa/2$ implies that
optimal ground state cooling leads to NMS only when the thermal noise
{[}first term in Eq.~(\ref{eqn:nopt})] is at least comparable to
the quantum backaction noise {[}third term in Eq.~(\ref{eqn:nopt})]. This is
likely to be the case in current endeavors to reach the ground state.\\
\indent \textit{Experimental realization}: To demonstrate that the observation
of parametric NMS is within experimental reach, we discuss the parameters
from Ref.~\citep{schliesser_resolved-sideband_2008}: $\Omega_{m}/2\pi=73.5\,\unit{MHz}$,
$\kappa/2\pi=3.2\,\unit{MHz}$, $\Gamma_{m}/2\pi=1.3\,\unit{kHz}$.
The cooling rate $\Gamma_{c}/2\pi=1.56\,\unit{MHz}$ was extracted
from the displacement spectrum's FWHM. A comparison with Eqs.~(\ref{q:shift})
then yields a coupling rate $g_{m}/2\pi\approx2.0\,\unit{MHz}$. Therefore,
the observation of parametric NMS is within experimental reach. In the electromechanical domain, using a superconducting
coplanar waveguide resonator, Ref.
\citep{teufel_dynamical_2008} reports coupling rates of $g_{m}/2\pi=6\,\unit{kHz}$
for a cavity with a decay rate $\kappa/2\pi=230\,\unit{kHz}$. \newline
In summary, we have analyzed a novel instance of NMS that occurs in
cavity optomechanics due to the coupling between the fluctuations
of the cavity field and the mechanical oscillator mode. Furthermore,
we have elucidated its implications for ground state cooling, namely,
the limitation through the cavity bandwidth.

We thank W. Zwerger for discussions and suggestions.
This work was funded by a Marie Curie Excellence Grant (RG-UHQ)and the DFG (NIM Initiative).

\end{document}